\documentclass[aps,prb,groupedaddress,twocolumn]{revtex4}
\usepackage{graphicx}
\include{epsf}
\begin{document}

\title{Resistive transition in $\pi$-junction superconductors}

\author{Enzo Granato}

\address{Laborat\'orio Associado de Sensores e Materiais, \\
Instituto Nacional de Pesquisas Espaciais, \\
12201-970 S\~ao Jos\~e dos Campos, SP Brazil}


\begin{abstract}
The resistivity behavior of inhomogeneous superconductors with
random $\pi$ junctions, as in high-$T_c$ materials with d-wave
symmetry, is studied by numerical simulation of a
three-dimensional XY spin glass model. Above a concentration
threshold of antiferromagnetic couplings, a resistive transition
is found in the chiral-glass phase at finite temperatures and the
critical exponents are determined from dynamic scaling analysis.
The power-law exponent for the nonlinear contribution found in
recent resistivity measurements is determined by the dynamic
critical exponent of this transition.

\end{abstract}

\pacs{74.25.Fy; 74.81.Bd; 74.81.Fa; 75.10.Nr}
\maketitle

Inhomogeneous superconductors containing a random distribution of
$\pi$ junctions, as in high-$T_c$ superconducting materials with
d-wave symmetry, can display unusual frustration effects even in
zero external magnetic field \cite{sigrist}. A $\pi$ junction
leads to a phase shift of $\pi$ between superconducting regions,
and to half-flux quantum vortices on closed loops with an odd
number of these junctions \cite{kirtley}. Interesting ordering
effects are expected due to vortex interactions. Some of them have
already been directly imaged on specially prepared high-$T_c$
Josephson contacts \cite{hilgemkamp} and should also be relevant
for $\pi$ junctions in low-$T_c$ superconductors \cite{lowtc}.
Much attention has been devoted to the magnetic properties of
inhomogeneous superconductors arising from the orbital currents of
theses vortices \cite{sigrist,kusmartsev,nordland} and in
particular to its relevance for the explanation of paramagnetic
Meissner effect \cite{braunisch}. Nevertheless, there are also
important consequences for the resistivity behavior of theses
systems which have not been investigated satisfactorily.

In the absence of $\pi$ junctions, the phases of neighboring
superconducting regions tend to be locked with zero phase shift,
and a phase-coherence transition is expected for decreasing
temperature into a superconducting state with vanishing linear
resistivity. The critical behavior of this resistive transition is
reasonable well understood. On the other hand, for sufficiently
large concentration of $\pi$ junctions, which may occur for
example in granular samples, frustration and disorder effects
leads to a vortex glassy phase and the resistive behavior is much
less understood. The simplest model of the system is to consider
only contributions from the Josephson coupling energy of
nearest-neighbor grains \cite{sigrist}, $H_{ij}=-J_o\cos(\theta_i
-\theta_j-t_{ij})$, where $\theta_i$ is the phase of the local
superconducting order parameter, $J_o > 0$ and $t_{ij}=0$ or $\pi$
correspond to the phase shifts of conventional and $\pi$
junctions. This is equivalent to the interaction of two-component
pseudo-spins $\vec{S}=(\cos(\theta),\sin(\theta))$, coupled by
ferro or antiferromagnetic interactions, respectively, which leads
to an XY-spin (chiral) glass model for the granular system
\cite{kawamura}. The chiral variable can be defined as the
direction of the local circulating currents (vortices) in closed
loops of junctions. Based on earlier and recent Monte Carlo (MC)
simulations \cite{kawamura,kawa97,kawa01} for this model in three
dimensions, it has been suggested that the equilibrium
low-temperature state for the inhomogeneous superconductor is a
chiral glass but with no phase coherence  and, therefore, the
resistivity should  be nonzero. This implies a chiral glass
transition  at a nonzero critical temperature but no resistive
transition, except perhaps at zero temperature. Thus, strictly
speaking, there is no true superconducting phase at low
temperatures in this scenario. However, while different works
agree on the existence of the proposed chiral glass transition,
the situation regarding the resistive behavior is unsettled.
Results for the ground state \cite{grempel,kost} of the XY-spin
glass model indicate that the lower critical dimension for phase
ordering is between $2$ and $3$ and therefore a phase-coherence
transition is only possible at zero temperature in two dimensions
\cite{eg} but should occur at finite temperatures in three
dimensions. The critical temperature for three dimensions,
however, can not be estimated from these calculations. Moreover,
dynamical simulations suggest a resistive transition at finite
temperature \cite{wengel,eg01} and very recent MC calculations for
a model with Gaussian couplings, expected to be in the same
universality class, strongly support the occurrence of this
transition \cite{leeyoung}. The dynamical simulations were based
on different representations of the same model and different
dynamics. While the static exponents agree, as expected from the
universality of critical behavior, the dynamic exponent $z \sim
4.6$ obtained from the resistively-shunted-junction (RSJ) model of
the dynamics in the phase representation \cite{eg01} is
significantly different from that obtained from MC dynamics, $z
\sim 3.1$, in the vortex representation \cite{wengel}, suggesting
a strong dependence of $z$ on the details of the dynamics.

On the experimental side, there have been some attempts to
identify  the chiral glass phase from nonlinear resistivity
measurements in ceramic $Y Ba_2 Cu_4 O_8$ bulk samples
\cite{yamao} at zero magnetic field, near the onset of the
paramagnetic Meissner effect. The nonlinear contribution $\rho_2$
to the resistivity was found to have a peak at the transition with
power-law behavior $\rho_2 \propto J^{-\alpha}$. This behavior has
already been reproduced in dynamical simulations \cite{dli}. The
results of the experiment have been interpreted as a chiral glass
transition attributed to the presence of $\pi$ junctions, with a
nonzero linear resistivity below the critical temperature, but the
value of $\alpha$ and its possible relation with the critical
exponents of the underlying transition was not found.

In this work, we study the resistivity behavior of inhomogeneous
$\pi$ junctions superconductors using an XY-spin glass model with
varying concentration $x$ of antiferromagnetic bonds. An improved
numerical method is used, combining MC and Langevin simulation
with periodic boundary conditions. The results of a scaling
analysis of extensive simulations for $x=0.5$ clearly show the
existence of a resistive transition at finite temperature. A
threshold $x_g \sim 0.3$ for the chiral glass phase is estimated
from the behavior of the zero-temperature critical current. The
power-law exponent $\alpha$ for the nonlinear contribution found
in resistivity measurements \cite{yamao} can be related to dynamic
critical exponent $z$ of this transition. The observed value of
$\alpha$ is within the range expected from numerical estimates of
$z$.

We consider inhomogeneous superconductors with $\pi$ junctions
modelled by a three-dimensional XY-spin glass described by the
Hamiltonian
\begin{equation}
H=-J_o\sum_{<ij>}\cos(\theta_i - \theta_j-t_{ij})  \label{xyspin}
\end{equation}
where the quenched phase shift $t_{ij}$ is equal to $\pi$ or $0$
with probabilities $x$ and $1-x$, respectively. The symmetric $\pm
J_o$ XY spin glass \cite{kawamura,grempel,eg,eg01,wengel}
corresponds to $x=0.5$ while the unfrustrated XY model correspond
to $x=0$. We use the time-dependent Guinzburg-Landau model for the
dynamics given by the Langevin equations
\begin{equation}
\frac{1}{R_o} \frac{d\theta_i}{dt} = - J_o\sum_j
\sin(\theta_i-\theta_j-t_{ij})+\eta_{i}            \label{dyn}
\end{equation}
where $\eta_{i}$ represents uncorrelated thermal noise with
$<\eta_{i}(t)\eta_i(t')>=2k_B T \delta(t-t')/R_o$ to ensure
thermal equilibrium. This can also be regarded as an onsite
dissipation model for the dynamics of the granular superconductor
where $R_o$ is the resistance of each point grain to the ground
and $J_o$ is the Josephson coupling. The RSJ model studied
previously \cite{eg01} allows only dissipation through the
junction shunt resistance. We use units where $\hbar/2e=1$,
$R_o=1$, $J_o=1$. To obtain the current-voltage characteristics
more accurately in the glassy phase, we introduce an improved
method. First, MC simulations are performed using (\ref{xyspin})
to obtain the equilibrium state (zero current bias) which is then
used as initial state to integrate numerically the Langevin
equations (\ref{dyn}) for the driven system. Periodic (fluctuating
twist) boundary conditions are used both for the MC simulations
\cite{benakli,egbjp} and driven Langevin dynamics \cite{dd99}
simulations. Previous simulations used current injection with free
boundary conditions \cite{eg01} but periodic boundary conditions
are more adequate since they avoid possible edge contributions.
For systems of linear size $L$, the voltage $V$ (electric field
$E=V/L$) was computed as a function of the driving current $I$
(current density $J=I/L^2$) for different temperatures and systems
sizes ranging from $L=4$ to $L=12$. Calculations were performed in
a cubic system, using  $10^7$ time steps and $10$ different
realizations of the $A_{ij}$ distribution, in the lowest current
range. The most extensive simulations were done for $x=0.5$ while
for $x < 0.5$ the main purpose was to obtain the qualitative phase
diagram and $T=0$ critical currents.

\begin{figure}
\includegraphics[bb= 1cm  2.5cm  19cm   26.5cm, width=7.5 cm]{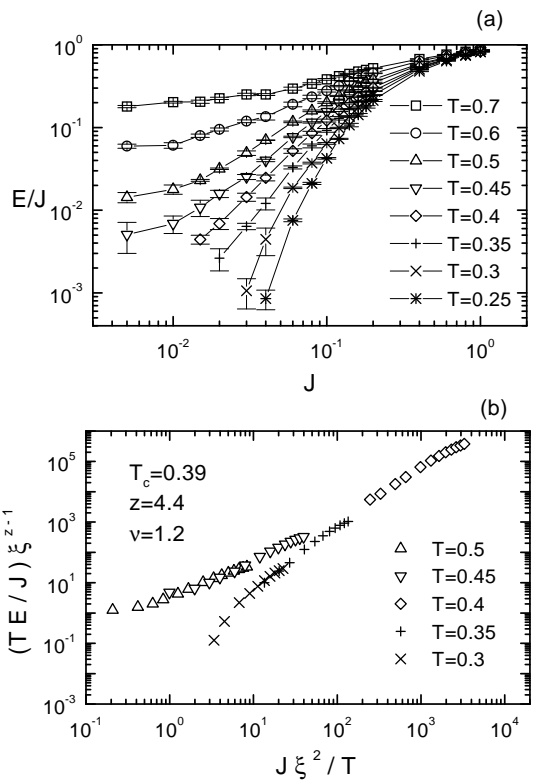}
\caption{(a) Nonlinear resistivity $E/J$ for $x=0.5$ ( equal
distributions of $0$ and $\pi$ junctions), and different
temperatures $T$, for system size $L=12$ ; (b) Scaling plot of the
data near the transition and for small currents, with $\xi \propto
|T/T_c - 1|^{-\nu}$. } \label{iv}
\end{figure}

The nonlinear resistivity $\rho=E/J$ for $x=0.5$ is shown in Fig.
1a for different temperatures $T$, for the largest system size
$L=12$. The behavior is consisting with a resistive transition at
an apparent critical temperature in the range $T_c \sim 0.3
-0.45$. At higher $T$, the linear resistivity
$\rho_L=\lim_{J\rightarrow 0} E/J$ is finite while  at lower $T$,
it extrapolates to zero. The phase transition can be confirmed by
a scaling analysis of the nonlinear resistivity which assumes the
existence of a continuous equilibrium transition at $T > 0$
\cite{fisher}. Near the transition, measurable quantities scale
with the diverging correlation length $\xi\propto |T-T_c|^{-\nu}$
and relaxation time $\tau \propto \xi^z$, where $\nu$ and $z$ are
the correlation-length and dynamical critical exponents,
respectively. The nonlinear resistivity should then satisfy the
scaling form \cite{fisher}
\begin{equation}
 T E \xi^{z-1}/J= g_\pm(J\xi^{2}/T)  \label{scaltc}
 \end{equation}
in $d=3$ dimensions where $g(x)$ is a scaling function. The $+$
and $-$ signs correspond to $T > T_c$ and $T < T_c$, respectively.
A scaling plot according to this equation can then be used to
verify the scaling arguments and the assumption of an underlying
equilibrium transition at $J=0$. The optimal data collapse
provides an estimate of $T_c$ and critical exponents. Such scaling
plot, which neglects finite-size effects, is shown in Fig. 1b,
obtained by adjusting the unknown parameters, giving the estimates
$T_c=0.39(2)$, $z=4.4(3)$ and $\nu = 1.2(2)$. We now show that
these estimates, using the largest system size, are reliable by
verifying that they give the expected finite-size behavior using
smaller system sizes. Finite-size effects are particularly
important sufficiently close to $T_c$ when the correlation length
$\xi$ approaches the system size $L$. In particular, at $T_c$, the
correlation length will be cut off by the system size in any
finite system and the nonlinear resistivity should then satisfy a
scaling form as in  Eq. \ref{scaltc} with $\xi = L$. In fact, as
shown in Fig. 2a, the nonlinear resistivity calculated at the
estimated $T_c=0.39$ for different system sizes satisfy this
scaling form with $z=4.6$ which agrees within the errors. Away
from $T_c$, the scaling function in Eq. \ref{scaltc} will also
depend on the dimensionless ratio \cite{fisher,wengel} $L/\xi$ as
$g(J\xi^{2}/T,L/\xi)$. To simplify the analysis, we consider
resistivity data at current densities such that $J\xi^{2}/T=$ is
constant. Then, the scaling form depends only on a single variable
and the resistivity should satisfy the finite-size scaling form
\begin{equation}
 T E L^{z-1}/J= \tilde{g}(L^{1/\nu}(T/T_c-1))  \label{scalL}
 \end{equation}
As shown in Fig. 2b, the nonlinear resistivity calculated for
different temperatures and system sizes such that $J\xi^{2}/T=1$,
with the estimated $T_c=0.39$ and $\nu=1.2$, indeed satisfy this
scaling form with $z=4.65$ which again agrees within the errors.
Using a different constant, $J\xi^{2}/T=2$, gives similar results.

\begin{figure}
\includegraphics[bb= 2cm  9.5cm  19cm   18.cm, width=7.5 cm]{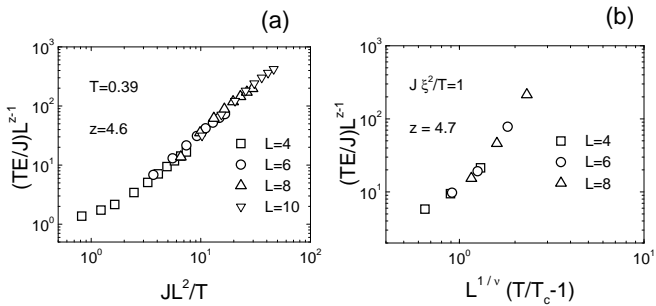}
\caption{(a) Finite-size scaling plot of the nonlinear resistivity
at $T_c=0.39$; (b) Finite-size scaling plot near $T_c$ using
current densities such that $J\xi^{2}/T=1$, a constant value.}
\label{ivtc}
\end{figure}

The values of $T_c$, $z$ and $\nu$ obtained by the above scaling
analysis using the onsite dynamics of Eq. \ref{dyn} agree well
with the previous estimate using the RSJ model \cite{eg01} for the
dynamics ($T_c=0.41(3)$, $z=4.6(4)$ and $\nu = 1.2(4)$ ), clearly
showing the existence of a phase-coherence transition at $T >0$
and also showing that the dynamic exponent $z$ is essentially the
same. Our estimate of $T_c$ from the resistivity scaling is in
good agreement with recent estimate of the critical temperature
for the chiral-glass transition from MC simulations \cite{kawa01},
in the range $T_{ch} =0.38 - 0.41$. The agreement is quite
intriguing since it supports the suggestion \cite{eg01} that
chirality and phase variables may order simultaneously. Recent MC
simulations of the XY-spin glass model with Gaussian couplings,
expected to be in the same universality class, strongly support
such single transition scenario \cite{leeyoung}. Nevertheless,
this transition is in sharp contrast with MC simulations of the
phase-overlap distribution function \cite{kawamura,kawa97,kawa01}.
On the other hand, a phase-coherence transition at $T>0$, is
consistent with calculations of the spin stiffness exponent in the
ground state showing that the lower-critical dimension for spin
order in the XY-spin glass model \cite{grempel} is below $3$ which
implies that a phase-coherence transition at $T > 0$ is possible.
More recently, improved calculations in the vortex representation,
also clearly shows a well-defined positive stiffness exponent
\cite{kost}. In addition, calculations of the linear resistivity
$\rho_L$ (zero current bias) from MC dynamics simulations in the
vortex representation \cite{wengel}, shows an equilibrium
resistive transition. The estimate of the static exponent $\nu$
agrees with the present estimate from the nonlinear resistivity
but the dynamic exponent \cite{wengel} $z=3.1$ is significantly
lower. Interestingly enough, our calculations of $z$ show the same
result for the onsite and RSJ dynamics. Additional calculations
using MC dynamics in the phase representation give the same result
\cite{egbjp}. In spite of that, it is possible that the different
$z$ is a result of the particular dynamics in the vortex
representation. In fact, vortex variables are collective
excitations in the phase representation and thus lead to
long-range correlations for the phases, suggesting that these
representations may belong to different dynamic universality
classes.

\begin{figure}
\includegraphics[bb= 2cm  9cm  19cm   17.5cm, width=7.5 cm]{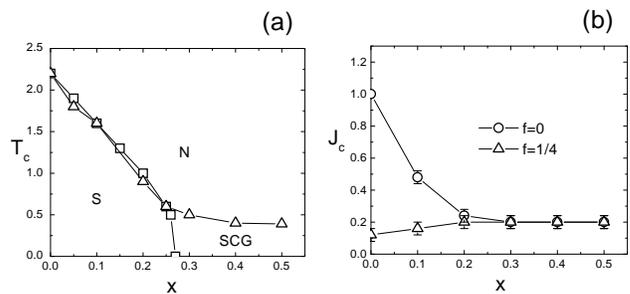}
\caption{(a) Phase diagram showing S (superconducting), SCG
(superconducting chiral glass) and N (normal) phases as a function
of temperature and concentration $x$ of $\pi$ junctions. (b)
Critical currents densities $J_c$ as a function of $x$ without
($f=0$) and with ($f=1/4$) a uniform external magnetic field. In
(a) triangle symbols correspond to critical temperatures estimated
from the resistivity behavior and squares from the
phase-susceptibility peak. } \label{phadia}
\end{figure}

The dependence of $T_c$ on the concentration of $\pi$ junctions
$x$ is shown in the phase diagram of Fig. 3a. The values of $T_c$
for $x < 0.5$ were obtained as rough estimates from the nonlinear
resistive behavior, and is found to be nonzero in the whole range.
We have also estimated the critical temperature from the peak of
the phase susceptibility $ \chi =(<m^2>-<m>^2)/L^3, $ where $m= |
\sum_i \vec{S}_i|$ and $\vec{S}=(\cos(\theta),\sin(\theta)$,
averaged over the disorder, which measures the onset of long-range
phase coherence outside the glassy phase. As shown in Fig. 3a,
this transition temperature decreases for increasing $x$ and
extrapolates to zero at a threshold value $x_g \sim 0.3$. We then
expect that the range $x> x_g$ should correspond to the vortex
(chiral) glass phase. We note that for $ x << x_g$ the
susceptibility peak agrees with $T_c$ showing that indeed the
resistive transition corresponds to the phase-coherence
transition. Additional evidence for the vortex glass phase is also
provided by the change of the critical current with applied
external field. Fig. 3b compares the behavior of the $T = 0$
critical current $J_c$ with and without a magnetic field $B$
applied transversely to the current direction. The external field
acts as a uniform frustration $f = Ba^2/\phi_o$ in the XY-spin
glass model of Eq. \ref{xyspin}, where $a$ is the lattice spacing
of the Josephson network and $\phi_o$ the flux quantum, and
introduces a vortex lattice with dimensionless spacing $a_v/a
\propto 1/f^{1/2}$. There is a large change of $J_c$ for $x <
x_g$, where some translational order at length scales large than
$a_v$ is still possible, but there is essentially no change for
$x> x_g$, indicating that in this range there is only short-range
order. The change of behavior gives a very rough estimate
\cite{dilu} of $x_g$.

Finally, we compare the critical properties of the resistive
transition with experiments. Nonlinear resistivity measurements in
ceramic $Y Ba_2 Cu_4 O_8$ bulk samples \cite{yamao} near the onset
of the paramagnetic Meissner effect have been interpreted as a
chiral glass transition attributed to the presence of $\pi$
junctions, with a nonzero linear resistivity below $T_c$. In the
experiments, the measured resistivity $\rho$ was separated into a
linear and nonlinear contribution through a low order expansion in
the current density, $\rho = \rho_o + \rho_2 J^2 + ...$. The
lowest order nonlinear contribution $\rho_2$ was found to have a
peak at the transition with power-law behavior $\rho_2 \propto
J^{-\alpha}$ and exponent $\alpha \sim 1.1(6)$, while the linear
contribution $\rho_o$ appears to remain finite below this
temperature. However, since the apparent linear contribution is
very small and finite current bias was used, the limited accuracy
of the data can not completely rule out a strict zero resistivity
phase below this temperature. It is of interest to verify to which
extent the resistive transition as found here is consistent with
the observed peak in the nonlinear resistivity at the apparent
transition temperature. If a resistive transition is assumed to
occur at this temperature then the power-law behavior of $\rho_2$
follows directly from the current-voltage scaling near the
transition temperature. Defining the nonlinear contribution as
$\rho_2 = \frac{\partial^2}{\partial J^2}(E/J) $, the scaling
behavior of Eq. (\ref{scaltc}), obeyed by our numerical data,
implies that $\rho_2 \propto J^{-\alpha}$, when $\xi \rightarrow
\infty$ near the critical temperature, with the exponent relation
$ \alpha = (5-z)/2 . $ Using the dynamical exponent of the
resistivity scaling \cite{zchiral}, $z=4.4(4)$ from the present
work and $z=3.1$ from the vortex-representation \cite{wengel},
gives the estimates $\alpha = 0.3(3)$ and $\alpha = 0.95$. These
are comparable to the observable value in the experiments within
the errors.

We should note that the model considered here neglects screening
of vortices due to inductance effects. For strong screening, the
finite-temperature transition is destroyed \cite{kawa97,wengel}.
However, it is possible that for very weak screening a resistive
transition is still possible.


\smallskip
This work was supported by FAPESP(Grant No. 03/00541-0).


\begin{thebibliography}{00}
\bibitem{zchiral} If the observed resistivity scaling just
reflects the critical singularity of the chiral order parameter,
as proposed in Ref. \cite{kawa01}, then, using the corresponding
dynamic exponent obtained from MC dynamics \cite{kawa01}, $z_{ch}
\sim 7.4(10)$, gives $ \alpha < 0$ which is inconsistent with the
experimental observation \cite{yamao} and numerical simulation
\cite{dli}.
\end{thebibliography}
\end{document}